\documentclass[trackchanges,twocolumn]{aastex631}

\newcommand{\her}{{\it Herschel}}

\newcommand{\arii}{\hbox{[Ar$\,${\scriptsize II}]}}
\newcommand{\ariii}{\hbox{[Ar$\,${\scriptsize III}]}}

\newcommand{\neii}{\hbox{[Ne$\,${\scriptsize II}]}}
\newcommand{\neiii}{\hbox{[Ne$\,${\scriptsize III}]}}

\newcommand{\siii}{\hbox{[S$\,${\scriptsize III}]}}
\newcommand{\siv}{\hbox{[S$\,${\scriptsize IV}]}}

\newcommand{\kms}{km\,s$^{-1}$} 
 
\newcommand{\um}{$\mu$m}

\newcommand\jwst{\emph{JWST}}

\newcommand\ifsfit{\texttt{IFSFIT}}

\newcommand\qtdfit{\texttt{q3dfit}}

\newcommand\fzeroeight{F08572$+$3915~NW}

\shorttitle{Warm H$_2$ Outflow in \fzeroeight}

\begin{document}

\title{\jwst\ Discovery of a Very Fast Biconical Outflow of Warm Molecular Gas in the Nearby Ultra-Luminous Infrared Galaxy \fzeroeight}

\author[0000-0001-5894-4651]{Kylie Yui Dan}
\affiliation{Department of Astronomy, University of Maryland, College Park, MD 20742, USA}

\author[0000-0002-4014-9067]{Jerome Seebeck}
\affiliation{Department of Astronomy, University of Maryland, College Park, MD 20742, USA}

\author[0000-0002-3158-6820]{Sylvain Veilleux}
\affiliation{Department of Astronomy, University of Maryland, College Park, MD 20742, USA}
\affiliation{Joint Space-Science Institute, Department of Astronomy, University of Maryland, College Park, MD 20742, USA}

\author[0000-0002-1608-7564]{David Rupke}
\affiliation{Department of Physics, Rhodes College, Memphis, TN 38112, USA}

\author[0000-0001-5285-8517]{Eduardo Gonzalez-Alfonso}
\affiliation{Universidad de Alcalá, Departamento de Física y Matemáticas, Campus Universitario, 28871 Alcalá de Henares, Madrid, Spain}

\author[0000-0002-9627-5281]{Ismael Garcia-Bernete}
\affiliation{Centro de Astrobiolog\'ia (CAB), CSIC-INTA, Camino Bajo del 497 Castillo s/n, E-28692 Villanueva de la Ca\~nada, Madrid, Spain}

\author[0000-0003-3762-7344]{Weizhe Liu}
\affiliation{Department of Astronomy, Steward Observatory, University of Arizona, Tucson, AZ 85719, USA}

\author[0000-0003-0291-9582]{Dieter Lutz}
\affiliation{Max Planck Institute for Extraterrestrial Physics, Giessenbachstraße 1, 85748 Garching, Germany}

\author[0000-0001-8485-0325]{Marcio Melendez}
\affiliation{Space Telescope Science Institute, 3700 San Martin Drive, Baltimore, MD 21218, USA}

\author[0000-0002-4005-9619]{Miguel Pereira Santaella}
\affiliation{Instituto de Física Fundamental (IFF), CSIC, Serrano 123, E-28006 Madrid, Spain}

\author[0000-0002-0018-3666]{Eckhard Sturm}
\affiliation{Max Planck Institute for Extraterrestrial Physics, Giessenbachstraße 1, 85748 Garching, Germany}

\author[0000-0002-6562-8654]{Francesco Tombesi}
\affiliation{Physics Department, Tor Vergata University of Rome, Via della Ricerca Scientifica 1, 00133 Rome, Italy}
\affiliation{INAF – Astronomical Observatory of Rome, Via Frascati 33, 00040 Monte Porzio Catone, Italy}
\affiliation{INFN - Rome Tor Vergata, Via della Ricerca Scientifica 1, 00133 Rome, Italy}

\begin{abstract}
We present new {\em James Webb Space Telescope} (\jwst) Mid-Infrared Instrument (MIRI) Medium-Resolution Spectrometer (MRS) observations of the nearby 
ultra-luminous infrared galaxy (ULIRG) \fzeroeight. 
These integral field spectroscopic (IFS) data reveal a kpc-scale warm-molecular rotating disk and biconical outflow traced by the H$_2$ $\nu$ = 0$-$0 S(1), S(2), S(3), and S(5) rotational transitions. The outflow maintains a relatively constant median (maximum) projected velocity of 1100 \kms\ (3000 \kms) out to $\sim$ 1.4 kpc from the nucleus. The outflowing H$_2$ material is slightly warmer (640 $-$ 700 K) than the rotating disk material (460 $-$ 520 K), perhaps due to shock heating in the highly turbulent outflowing material. This outflow 
shares the same kinematics and orientation as the 
sub-kpc scale warm-H$_2$ outflow traced by the ro-vibrational H$_2$ lines in Keck AO near-infrared IFS data. However, this warm-H$_2$ outflow is significantly faster than the sub-kpc scale cold-molecular outflow derived from multi-transition far-infrared OH observations with \her\ and the $\ga$ kpc-scale cold-molecular outflow mapped by mm-wave interferometric CO 1$-$0 observations with IRAM-PdBI and NOEMA. The new \jwst\ data bolster the scenario where the buried quasar in this ULIRG is excavating the dust screen, accelerating perhaps as much as 60\% of the dusty warm-molecular material to velocities beyond the escape velocity, and thus influencing the evolution of the host galaxy.

\end{abstract}


\section{Introduction} \label{sec:intro}
Our current picture of galaxy evolution requires some form of feedback mechanism to regulate star formation. Models without feedback consistently over-predict the abundance of galaxies at the highest and lowest mass ends when compared to observations. At the low mass end, stellar feedback is thought to regulate the smallest galaxies \citep[e.g.,][]{dekel1986,somerville1999,Benson2003,dekel2003}. At the high mass end, feedback from active galactic nuclei (AGN) is purported to play a role in regulating star formation \citep[e.g.,][]{cole2000, Benson2003, croton2006, sijacki2007}, but how AGN do so is not yet completely understood. Additionally, any feedback mechanisms must be able to explain the properties of the quiescent (``red-and-dead"), massive ellipticals which make up the majority of the population of high-mass galaxies \citep[e.g.,][and references within]{veilleux2005, fabian2012, heckman2014, veilleux2020, heckman2023}. 

Galaxy mergers have proven to be great laboratories to study feedback. During a gas-rich merger, a large fraction of the gas  is funneled into the circumnuclear region of the merger remnant, leading to both a circumnuclear starburst and the growth of a supermassive black hole (SMBH) through accretion. As the SMBH grows, the energy released by the SMBH may quench the starburst through the expulsion of the interstellar medium (ISM) from which stars are formed. In the local universe, these systems are often identified as ultraluminous infrared galaxies (ULIRGs) with infrared (8 $-$ 1000 \um) luminosities above 10$^{12}$ $L_\odot$.  
Galaxy-wide outflows have been detected in many ULIRGs
\citep[e.g.,][and references within]{veilleux2020}. 

Outflows observed using molecular gas tracers are key to understanding this feedback process, not only because the molecular phase may carry a significant fraction of the outflow mass and momentum, but also because molecular gas is the raw material for star formation. Most of the research effort has so far focused on the cold-molecular phase of these outflows, but the successful deployment of \jwst\ is now allowing the study of the warm-molecular gas phase in exquisite detail. 

In this paper, we report the first results from a cycle 2 \jwst\ program to study  ULIRGs with known fast and powerful molecular outflows (PID 3869, PI Veilleux). The target is \fzeroeight, the northwestern component of the well-known, nearby ($z = 0.0576$), early-phase merger made up of two distinct galaxies separated by 5.7 kpc 
\citep{surace1998,kim2002, veilleux2002}. 
\fzeroeight\ is a ULIRG with a highly obscured AGN that contributes $\sim$ 70\% of the total bolometric luminosity of the system \citep[][]{veilleux2009}. High-velocity outflows have been detected in the neutral, ionized, and molecular gas phases.
Gemini Multi-Object Spectrograph integral field spectroscopic (IFS) observations reveal spatially resolved (kpc-scale) outflows of blueshifted ionized and neutral gas with maximum velocities of $-$3345 km s$^{-1}$ and $-$1153 km s$^{-1}$, respectively \citep{rupke2013_gemini}. 
Complementary Keck adaptive optics near-infrared IFS observations of the H$_2$ $\nu$ = 1$-$0 transitions trace a sub-kpc blueshifted outflow with maximum velocity of $-$1700 \kms\ \citep{rupke2013, rupke2016}. 
The nuclear ($<$ kpc) spectra from \her\ PACS far-infrared observations also display evidence of a fast (up to $\sim$ 1260 \kms) cold-molecular outflow based on the detection of deep blueshifted absorption troughs in the OH features \citep{sturm2011, veilleux2013, gonzalez-alfonso2017}.
Finally, IRAM PdBI and NOEMA CO (1$-$0) observations show a blueshifted outflow with maximum velocity of $-$1200 \kms\ and a redshifted outflow with maximum velocity of $+$1100 \kms\ within the inner $\sim$ 1.4 kpc, as well as a separate $\sim$ $+$900 \kms\ cloud located 6 kpc northwest of \fzeroeight\ that may also be outflowing (\citealt{cicone2014},  \citealt{herrera2020}). 

In this paper, we complement the multi-wavelength analysis of \fzeroeight\ with the IFS MIRI/MRS data on the warm molecular gas. In Section \ref{sec:analysis}, we report the observations and data analysis of the \jwst\ data. We present our results in Section \ref{sec:results} and discuss them in Section \ref{sec:discussion}.
We adopt a cosmology of $H_0 = 69.6$, $\Omega_M = 0.286$, and $\Omega_L = 0.714$, which implies a scale of 1.137 kpc arcsec$^{-1}$ and a luminosity distance of 262.6 Mpc.

\section{Observations and Data Analysis}\label{sec:analysis}

\fzeroeight\ was observed with \jwst\ on 2024-04-16 using the Medium-Resolution Spectrometer (MRS) mode of the Mid-InfraRed Instrument \citep[MIRI;][PID 3869, PI Veilleux]{wright2023,argyriou2023}. All of the grating settings -- Short, Medium, and Long -- were used to cover the entire wavelength range of MIRI/MRS (4.9 to 27.9 $\mu$m). A 4-point extended dither pattern was used to help remove background contamination and reduce undersampling, resulting in a total exposure time of 2697 s per channel. A background cube was obtained immediately after the on-source cube using the same settings, with the exception of using the 2-point dither, resulting in an exposure time of 1349 s per channel. We reduced these data with \jwst\ Calibration Pipeline v1.15.1 \citep{Bus2022, Bus2024} and v11.17.26 of the Calibration Reference Data System (CRDS) using the master-background subtraction setting. The data were combined with those from PID 3368 (PI Armus) to raise the signal-to-noise ratio. The data described here may be obtained from the MAST archive at \dataset[doi:10.17909/dbvp-0d64]{https://dx.doi.org/10.17909/dbvp-0d64}. Additionally, we corrected the WCS of each cube to make the location of the brightest spaxel in integrated light equal to the center of \fzeroeight. 

We fit the data spaxel-by-spaxel with the software package \qtdfit\ \citep[][]{q3d_rupke2023}, inspired from \ifsfit\ \citep{rupke2014}. 
\qtdfit\ is designed specifically for the analysis of IFS data on QSOs, as it removes the bright point spread function (PSF) caused by the central compact QSO emission, revealing the much fainter emission from the host galaxy without contamination from the bright PSF. We use \qtdfit\ to extract a spectrum to use as a PSF template using a 1-pixel radius around the brightest spaxel. For each spaxel, the template is scaled up or down using a second-order exponential to match the continuum and remove any signal that resembles the nuclear spectrum. Examples of spectra at two spaxels near the nucleus, before and after PSF subtraction, are shown in Figure \ref{fig:psf_comp}. The right panel has a ($\Delta$x,$\Delta$y) from the center of (-0.378, 0) kpc, while the left panel has (-0.378, -0.378) kpc.
More details on the \qtdfit\ procedure and its use in the analysis of \jwst \ IFS data can be found in \citet{Wylezalek2022,  Rupke2023, Veilleux2023, Vayner2023} 

Since the main focus of the present paper is the warm molecular gas, we fit H$_2$ $\nu$ = 0$-$0 S(1) 17.03$\mu$m, S(2) 12.27$\mu$m, S(3) 9.67$\mu$m, and S(5) 6.91$\mu$m in the data cube (S(4) 8.02$\mu$m is detected but very faint; it is not considered thereafter). For comparison, we also examine the fine structure lines \neii\ 12.81$\mu$m, \neiii\ 15.56$\mu$m, \siii\ 18.71$\mu$m, \siv\ 10.51$\mu$m, \arii\ 6.99$\mu$m, and \ariii\ 8.99$\mu$m, tracers of the warm ionized gas. We use one or two Gaussian components to fit the emission line profiles. We require the Gaussian components to be detected at the 3-$\sigma$ level, except in a few spaxels where a second component below the 3-$\sigma$ threshold is allowed after visual inspection of the data. 
\begin{figure}
    \centering
    \includegraphics[width=\columnwidth]{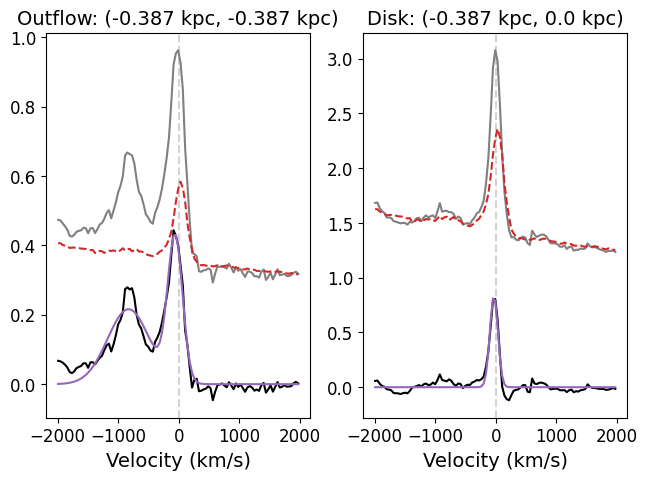}
    \caption{Examples of spectra at spaxels with ($\Delta$x,$\Delta$y) from the center of (-0.378, 0) kpc and (-0.378, -0.378) kpc for the right and left panels respectively, showcasing the H$_2$ 0$-$0 S(3) line profiles near the nucleus of \fzeroeight. Grey solid lines represent the data, red dashed lines show the PSF-scaled spectra, solid black lines display the PSF-subtracted spectra, and solid purple lines trace the line fits with either one (right) or two (left) Gaussian components. Through the PSF subtraction capabilities of \qtdfit, we can more readily distinguish between flux from the PSF and from the extended emission of the disk and outflow. \label{fig:psf_comp}}
\end{figure}

\begin{figure*}
    \centering
    \includegraphics[width=\textwidth]{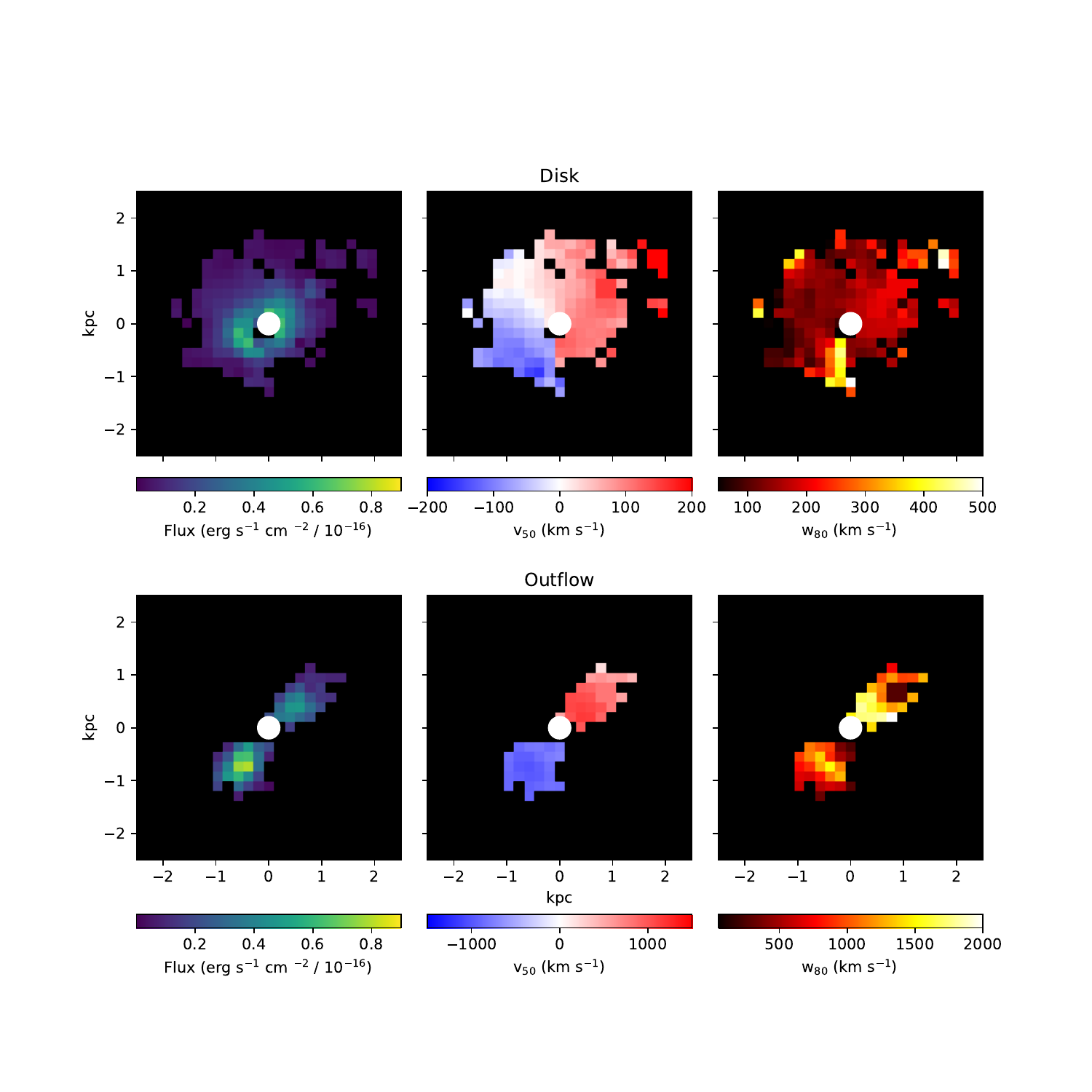}
    \caption{Results from the two-Gaussian fits to the H$_2$ 0$-$0 S(3) line profiles in \fzeroeight. In all panels, north is to the top and east to the left. The pixel scale is 0.17 arcsec px$^{-1}$. The white circle represents the spaxels utilized for PSF subtraction, where the fits are therefore unavailable. Top row: Maps of the line fluxes, median velocities (v$_{50}$), and 80-percentile line widths ($w_{80}$) of the narrow Gaussian component. The narrow component traces a quiescent (kinematically ``quiet") disk rotating with a maximum velocity amplitude of $\pm$150 \kms. Bottom row: Same as above but for the broad Gaussian component. This component traces a highly turbulent biconical outflow with $ v_{50}$ of up to $\pm$1100 km s$^{-1}$.
    \label{fig:disk_main}}
\end{figure*}

\section{Results}\label{sec:results}

The profiles of the H$_2$ rotational lines in \fzeroeight\ are often complex, requiring fits that involve two kinematically distinct Gaussian components. Figure \ref{fig:disk_main} shows the results of these fits for H$_2$ 0$-$0 S(3), the strongest of these transitions in the data. The flux map for each Gaussian component is shown in the left column of this figure; the median (50-percentile) velocities, $v_{50}$, in the middle column; and the 80-percentile line widths, $w_{80} \equiv | v_{90} - v_{10} |$, where $v_{90}$ and $v_{10}$ are respectively the 90- and 10-percentile velocities, in the right column. The narrow Gaussian component in the fits (top row in Figure \ref{fig:disk_main}) traces a ``kinematically quiet" ($w_{80} \la$ 300 \kms) and slightly off-centered, molecular gas disk in rotation around the nucleus with a maximum velocity amplitude of $\pm$150 \kms\ with respect to the systemic velocity ($z=0.0576$, directly estimated from the kinematics of H$_2$ disk). 
The few spaxels with $w_{80} > 300$ \kms\ reflect a degeneracy between the broad and narrow component fits. Both components are still fit, but the $w_{80}$ and $v_{50}$ values are less reliable. The broader Gaussian component (bottom row in Figure \ref{fig:disk_main}) reveals a fast ($\vert v_{50} \vert \sim$ 1000 \kms), kinematically turbulent ($w_{80} \sim$ 1000 $-$ 2000 \kms), biconical outflow with an opening angle of $\sim$ 50 $\pm$ 10$^\circ$ that extends up to $\sim$ 1.4 kpc from the nucleus along position angle P.A. $\approx$ 135 $\pm$ 15$^\circ$ (measured from North to East), roughly aligned with the kinematic major axis of the disk. The blueshifted side of the outflow is $\sim$ 2 $\times$ brighter than the redshifted side. The broad outflow component is detected in the nuclear region (within $\sim$ 0.25 kpc of the center), but as those spaxels are used for PSF subtraction, the fits are not shown in Figure \ref{fig:disk_main}.

\begin{figure*}
    \centering
    \includegraphics[width=\textwidth]{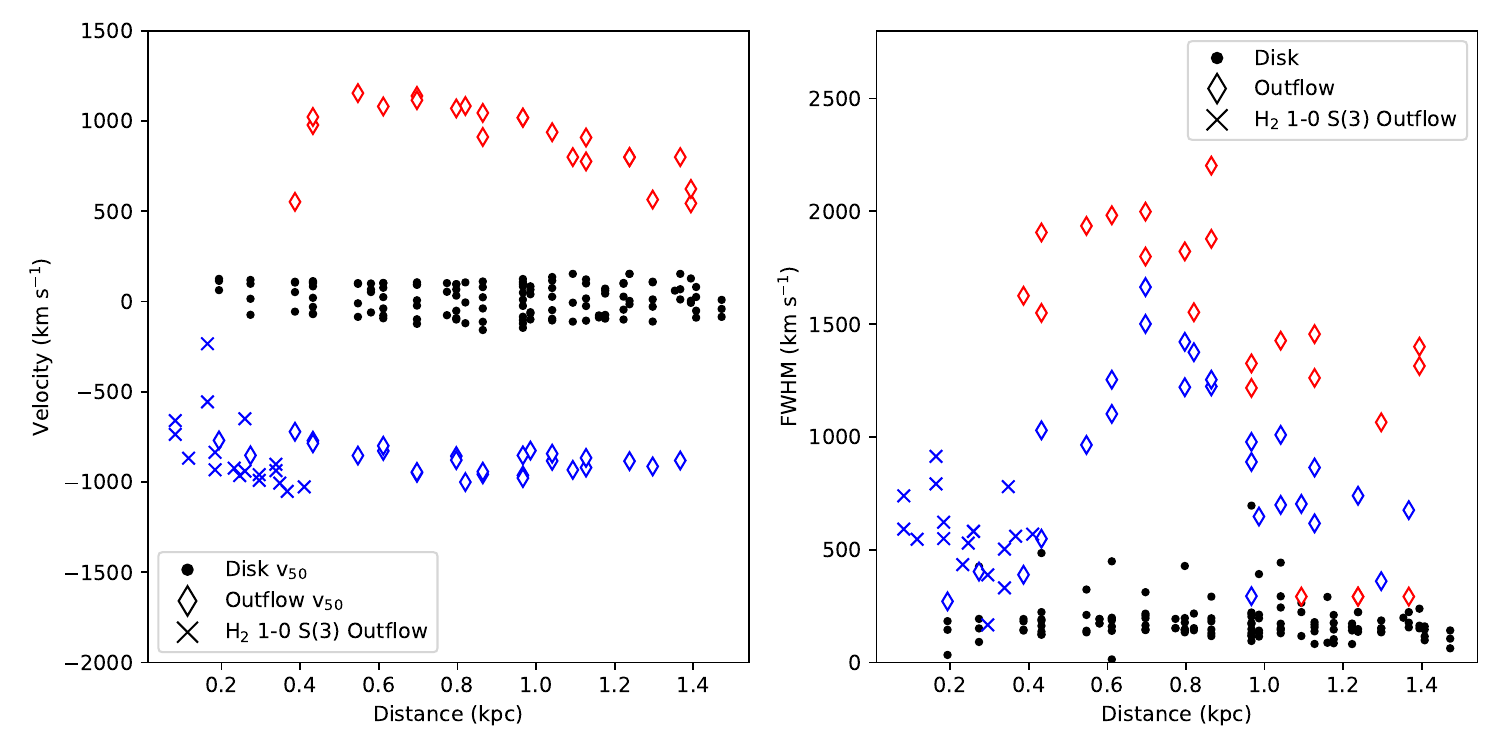}
    \caption{Kinematics of the warm-H$_2$ gas derived from H$_2$ 0$-$0 S(3). Left: Median velocities (v$_{50}$) as a function of projected distance from the galaxy center. Black circles represent the rotating disk component and diamonds represent the outflow component with blue diamonds indicating the blueshifted side and red diamonds indicating the redshifted side. For comparison, the blue X's mark the outflow velocities based on H$_2$ 1$-$0 S(3) from \citet{rupke2013}. Right: The full widths at half maximum (FWHM) of H$_2$ 0-0 S(3) as a function of projected distance from the center using the same symbols as the left panel. The estimated errors on $v_{50}$ and FWHM are typically the same size as the symbols (on the order of 100 \kms\ or less), so they are not shown in the figure. 
    \label{fig:rotation_curve}}
\end{figure*}

The kinematics of the warm molecular gas as a function of projected distance from the galaxy center are presented in Figure \ref{fig:rotation_curve} for the H$_2$ 0$-$0 S(3) transition. The black filled circles show the low-velocity disk, while the blue and red diamonds mark the high-velocity blueshifted and redshifted sides of the biconical outflow, respectively. For comparison, the results from \citet{rupke2013, rupke2016} based on the near-infrared H$_2$ 1$-$0 S(3) 1.96 $\mu$m transition are shown as blue X's in this figure. \citet{rupke2013, rupke2016} found that the $v_{50}$ of the blueshifted outflow in H$_2$ 1-0 S(3) grows from $-$700 \kms\ at 0.3 kpc from the center to $-$1000 \kms\ at 0.4 kpc. The values of $v_{50}$ for H$_2$ 0$-$0 S(3) continue this trend from 0.4 kpc to 0.6 kpc, reaching a peak value of $-$1100 \kms\ near $\sim$ 0.8 kpc. At distances beyond 0.8 kpc, the values of $\vert v_{50} \vert$ on the blueshifted side of the outflow decrease slightly but remain above 900 \kms\ out to 1.4 kpc. Similarly, the values of the full widths at half maximum (FWHM) on the blueshifted side increase to $\sim$ 1700 \kms\ out to $\sim$ 0.7 kpc and decrease to $\sim$ 600 $-$ 1000 \kms\ beyond 0.8 kpc, although these trends are uncertain given the larger scatter in the measurements of FWHM than for $v_{50}$. 

The kinematic behavior of the redshifted side of the outflow in H$_2$ 0$-$0 S(3) is remarkably similar to that of the blueshifted side, both in terms of $\vert v_{50} \vert$ and FWHM: $v_{50}$ increases from $+$500 \kms\ at 0.4 kpc from the center to $+$1100 \kms\ near 0.6 kpc. Beyond that point, $v_{50}$ gradually declines to $\sim$ 450 $-$ 900 \kms. This steeper decline in $v_{50}$ on the fainter redshifted side of the outflow may partially be due to the larger measurement uncertainties associated with distinguishing the faint redshifted outflow emission from the faint disk emission beyond $\sim$ 1 kpc from the center.

Figure \ref{fig:main} compares the emission line profiles of H$_2$ 0$-$0 S(1), S(3), and S(5) in the blueshifted and redshifted sides of the outflow (extracted from regions B and R, respectively, as defined in the middle panel of Figure \ref{fig:main}). The locations of regions B and R were selected to ensure detection of the outflow across all four H$_2$ 0$-$0 lines, taking into account the different PSF sizes across MIRI/MRS channels and the underlying noise structure. The characteristic line splitting detected in H$_2$ S(3) is also present in the other transitions. The narrow components in all cases lie within $\pm$120 \kms\ of the systemic velocity, while the broad component peaks near $\pm$1000 \kms. Figure \ref{fig:main} also shows that the strength of the broad outflow component relative to that of the narrow disk component increases with H$_2$ 0$-$0 excitation level: it is only weakly detected in the S(1) line while it is comparable in strength in S(3) and dominates the overall emission in S(5). This trend indicates a higher excitation temperature in the outflowing material than in the disk; we return to this point in Section \ref{sec:discussion} where we derive the excitation temperatures in both components. 

\begin{figure*}
    \centering
    \includegraphics[width=\textwidth]{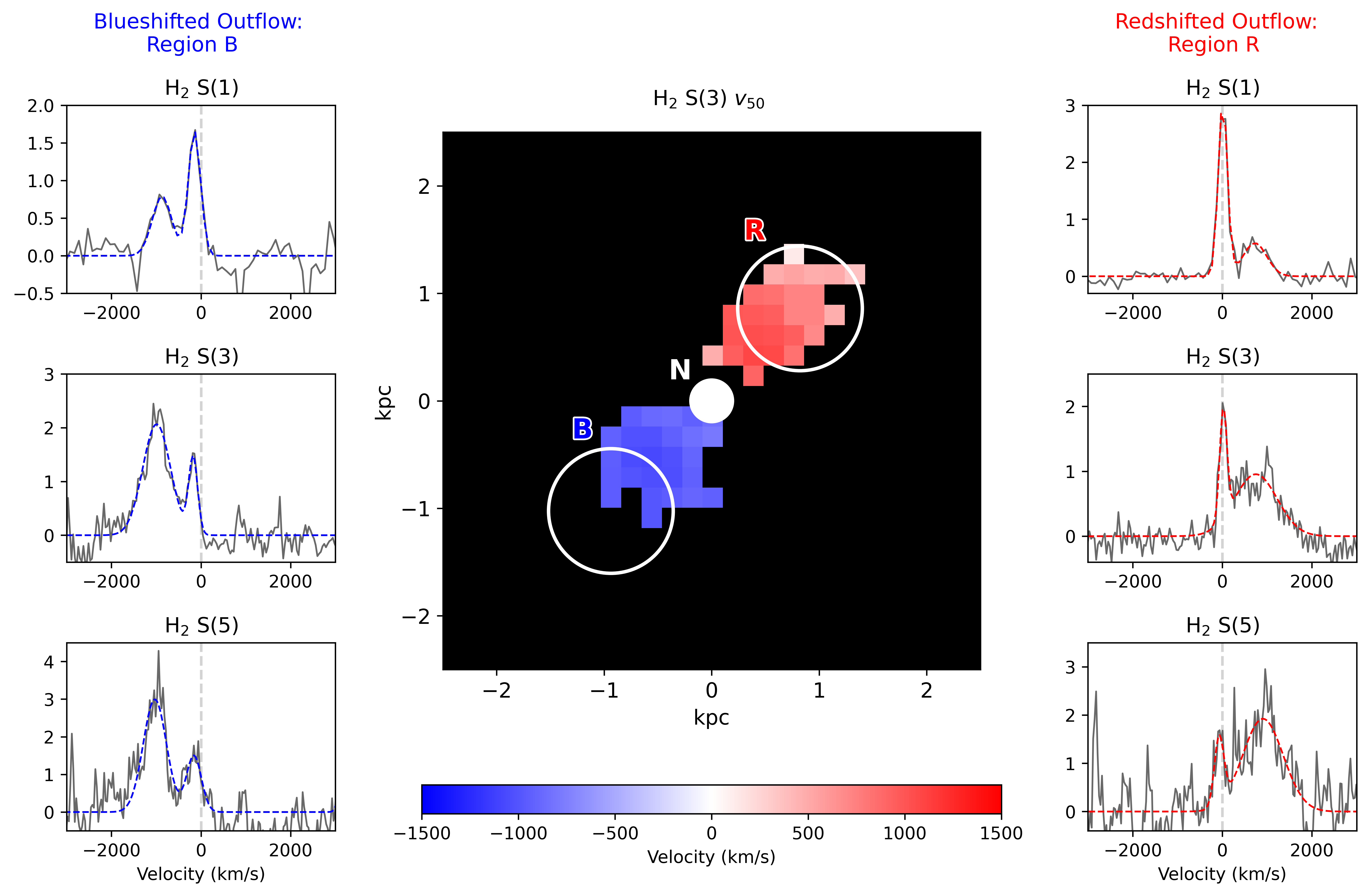}
    \caption{Comparisons of the line profiles of H$_2$ 0$-$0 S(1), S(3), and S(5) in the outflow (H$_2$ 0$-$0 S(2) is also detected but not shown here). The left panels show the line profiles extracted from region B, capturing the blueshifted side of the outflow, while the right panels show the line profiles from region R, capturing the redshifted side. The units of the vertical axes are $10^{-14}$ erg s$^{-1}$ cm$^{-2}$ $\mu$m$^{-1}$. The exact locations of regions B and R are shown in the central panel, superposed on the v$_{50}$ map of H$_2$ 0$-$0 S(3). 
    The grey solid lines in the left and right panels represent the data and the blue and red dashed lines represent the fits.  For all profiles, the fits require two Gaussian components: one narrow component representing the rotating disk and one broad component representing the outflow. Note the systematic change in the relative intensity of the disk and outflow components (see text for further details). 
    \label{fig:main}}
\end{figure*}

We do not find any clear evidence for the broad outflow component in any of the fine structure lines. This is illustrated in Figure \ref{fig:compare}, where we compare the high-S/N profiles of the low-ionization \neii\ 12.81$\mu$m line with those of H$_2$ 0$-$0 S(3) in the outflow regions. For completeness, we also show the profiles in the nuclear region, where the high-velocity outflow component is again detected in H$_2$ but not in \neii. The lack of a clear detection of the outflow component in \neii\ 12.81$\mu$m and the other fine structure lines indicates the warm ionized gas phase is negligible in this outflow. We return to this point in Section \ref{sec:discussion} below.

\begin{figure*}[ht!]
    \includegraphics[width=\textwidth]{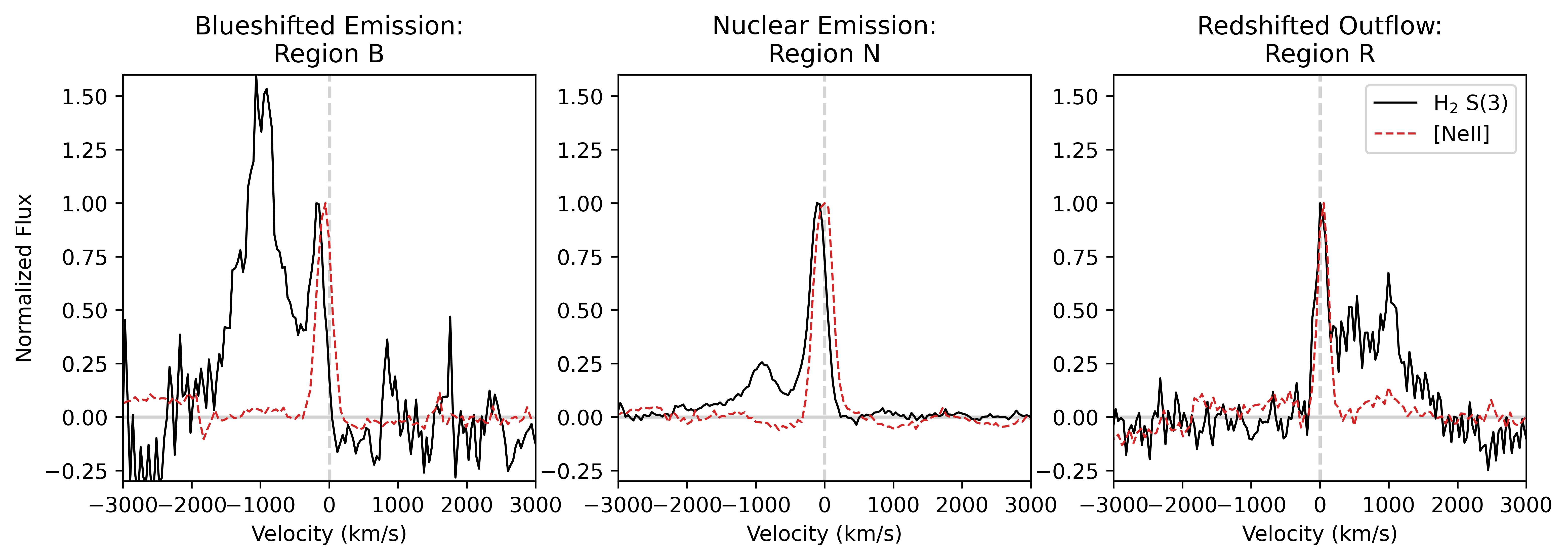}
    \caption{Comparison of the line profiles of H$_2$ 0$-$0 S(3) (black solid line) and \neii\ (red line) in regions B, N, and R, as defined in Figure \ref{fig:main}. The broad outflow component is not clearly detected in \neii. 
    \label{fig:compare}}
\end{figure*}

\section{Discussion}\label{sec:discussion}

The new MIRI/MRS IFS data on \fzeroeight\ have revealed the existence of a biconical (opening angle $\sim$ 50 $\pm$ 10$^\circ$) warm-molecular outflow with median absolute projected velocities $\vert v_{50} \vert \sim$ 1000 \kms\ and broad profiles with FWHM $\sim$ 1000 $-$ 2000 \kms\ out to 1.4 kpc along P.A.\ $\sim$ 135 $\pm$ 15$^\circ$. The redshifted side of the outflow is a nearly symmetric reflection of the blueshifted side in terms of kinematics, but it is $\sim$ 2 $\times$ fainter than the blueshifted side, likely attenuated by the intervening warm-molecular disk detected in the MIRI/MRS data. These results are remarkably consistent with the one-sided blueshifted conical outflow detected by \citet{rupke2013, rupke2016} in the near-infrared H$_2$ 1$-$0 emission lines, assuming the redshifted outflow in the near-infrared data is obscured by the intervening material in the host galaxy. Taken together, these data paint a big picture where the outflowing material is accelerating radially out to 0.8 kpc and then largely cruises at constant speed from 0.8 to 1.4 kpc with little evidence for significant deceleration. The broad line profiles throughout the outflow suggest that the entrained molecular material is highly turbulent and volume-filling rather than material that lies only on the outer surface of the bicone.  Despite the collimated nature of the H$_2$ 0-0 outflow, there is no evidence of a radio jet in this object: radio data at 1.4 GHz, 8.44 GHz, and 33 GHz show dominant compact emission with little to no extended emission \citep{condon1990, condon1991, barcos2017}. 

Overall, the velocities 
of the warm-molecular outflow in the \jwst\ data are significantly higher than those of the cold-molecular outflow deduced from the multi-transition OH absorption line profiles in the \her\ spectra \citep[$v_{\rm max}$ $\sim$ $-$1300 \kms;][]{sturm2011, veilleux2013, gonzalez-alfonso2017} and CO (1$-$0) interferometric maps \citep[$v_{\rm max}$ $\sim$ $-$1300 \kms;][]{cicone2014, herrera2020}.
To estimate the mass involved in the warm-molecular outflow and compare it with that of the cold-molecular outflow, we use the transitional data from \citet{roueff2019}
and Equation 2 of \citet{youngblood2018} to create an excitation diagram from the extinction-corrected fluxes of all detected H$_2$ lines (Figure \ref{fig:excite}). We assume an ortho-to-para ratio of 3:1 and negligible mid-infrared extinction outside of the nucleus (we return to these assumptions below). The H$_2$ temperature is inferred as the slope from a linear fit to the data, using Equation 2 from \citet{youngblood2018}:
\begin{equation}
\label{eqn:h2temp}
    \log_{10} \frac{N(v_u, J_u)}{g(J_u)}=-\frac{1}{T\cdot \ln(10)} \frac{E(v_u, J_u)}{k_B} + \log_{10} N(0,0), 
\end{equation}
where $N(v_u, J_u)$ is the level H$_2$ column density as described in Equation 1 of \citet{youngblood2018}, $g(J_u)$ is the level degeneracy, $T$ is the temperature, $\frac{E(v_u, J_u)}{k_B}$ is the upper-level energy, and $N(0,0)$ is the total H$_2$ column density $N_{H_2}$ scaled by the partition function $Z(T)$ from \cite{roussel2007}. Dashed lines in Figure \ref{fig:excite} represent the linear fits to the four H$_2$ 0$-$0 data points. The quality of these fits indicates that our assumptions on the ortho-to-para ratio and mid-infrared extinction are reasonable. In region B, the outflow temperature ($T = 640 \pm 90$ K) is not significantly different from the disk temperature ($T = 520 \pm 180$ K). In region R, the outflow temperature ($T = 700 \pm 180$ K) is significantly hotter than the disk temperature ($T = 460 \pm 150$ K). Heating due to shocks in the highly turbulent outflowing material may be at the origin of the higher H$_2$ temperature in the outflow components.

\begin{figure*}
    \centering
    \includegraphics[width=\textwidth]{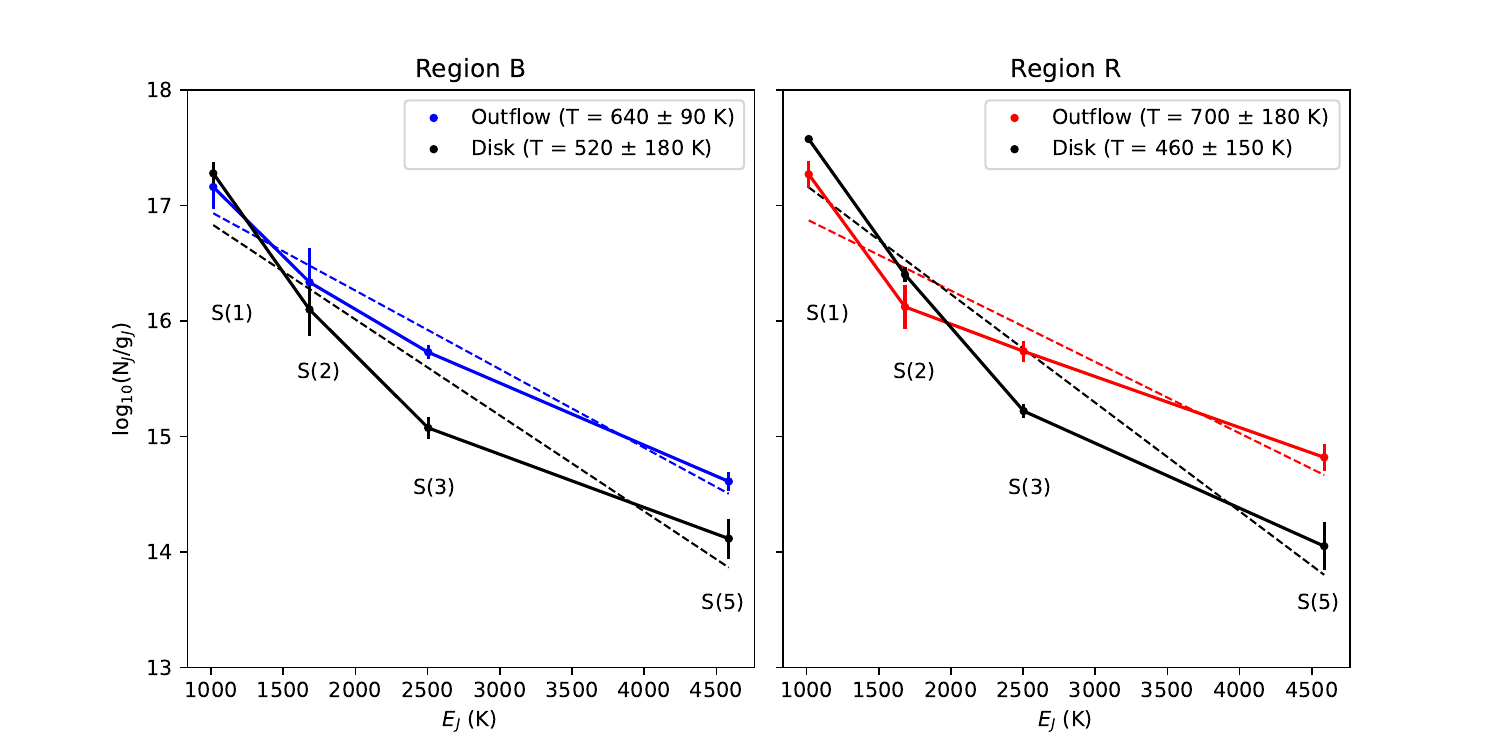}
    \caption{Excitation diagrams based on the H$_2$ 0$-$0 lines in region B (left panel) and region R (right panel). The colored points represent the outflow components, while the black points represent the disk component. The dashed lines are linear fits where the gas temperature is inferred from the slope and the H$_2$ column density from the y-intercept, as expressed by Equation \ref{eqn:h2temp}. 
    \label{fig:excite}}
\end{figure*}

Next, we use the H$_2$ column densities $N_{H_2}^{\rm blue} = 6.62 \times 10^{18}$ cm$^{-2}$ and $N_{H_2}^{\rm red} = 1.47 \times 10^{19}$ cm$^{-2}$, inferred from the y-intercepts of the linear fits to Equation \ref{eqn:h2temp} for both the blueshifted and redshifted sides of the outflow, to estimate the mass of warm H$_2$ in the entire outflow, $M_{\rm out}$: 
\begin{equation}
\label{eqn:mass}
    M_{\rm out} = M_{\rm out}^{\rm blue} + M_{\rm out}^{\rm red} = \frac{m_{H_2}}{M_\odot}\left(N_{H_2}^{\rm blue} A_{\rm out}^{blue} + N_{H_2}^{\rm red} A_{\rm out}^{\rm red}\right) , 
\end{equation}
where $\frac{m_{H_2}}{M_\odot}$ is the mass of an H$_2$ molecule in solar masses and $A_{\rm out}^{\rm blue}$ and $A_{\rm out}^{\rm red}$ are the areas covered by the blueshifted and redshifted sides of the outflow, respectively, determined from the H$_2$ 0-0 S(3) map. Here, we have assumed that the physical conditions ($T$, $N_{H_2}$) in regions B and R are representative of the entire blueshifted and redshifted outflows, respectively. We obtain $M^{\mathrm{blue}}_{\mathrm{out}} = 1.06 \times 10^5 \mathrm{\ M}_\odot$ for region B and $M^{\mathrm{red}}_{\mathrm{out}} = 8.77 \times 10^4 \mathrm{\ M}_\odot$ for region R, with a total outflow mass of $M_{\rm{out}} = 1.94 \times 10^5 \mathrm{\ M}_\odot$. 

Despite sharing similar kinematics with the cold-molecular, neutral-atomic, and warm-ionized gas phases of the outflow, the warm-molecular gas phase therefore respectively represents only $\sim$ 0.1\%, 0.3\%, and 2\% in mass relative to these other gas phases \citep{gonzalez-alfonso2017, herrera2020, rupke2013_gemini}. These are among the first measurements of these ratios in galactic outflows \citep[e.g.,][]{davies2024, seebeck2024}. The warm molecular gas probed by the H$_2$ 0-0 lines is thus only a tracer of the dominant cool outflowing gas. In comparison, the global warm-to-cold molecular gas ratios measured in starburst and Seyfert galaxies are 1$-$2 orders of magnitude larger than our value \citep[]{rigopoulou2002, pereira2014}. The non-detection of the outflow in the fine structure lines confirms the results of \citet{rupke2013_gemini} that the warm-ionized component of this outflow is also dynamically insignificant. 
We estimate an upper limit on this mass from our \neii\ spectra of $1.0\times 10^9$ M$_\odot$, which is about two orders of magnitude greater than the mass estimated by \citet{rupke2013_gemini}.

The outflow velocities of the warm and cold molecular gas in \fzeroeight\ are among the largest values ever measured in any galaxy, near and far. Maximum outflow velocities v$_{\mathrm{max}}$ based on the 90- or 10-percentile of the H$_2$ 0$-$0 S(3) line emission in the data reach values of order 3000 \kms. This number should be considered a lower limit to the actual maximum velocity, given likely projection effects. 
With an estimated escape velocity $v_{\mathrm{esc}} \approx 850$ km s$^{-1}$ for \fzeroeight\ \citep{herrera2020} and measured H$_2$ 0$-$0 outflow velocities remaining above $\sim 1000$ km s$^{-1}$ past 1 kpc, a significant fraction of the warm molecular gas is expected to be launched out of the galaxy. More precisely, we find that $\sim$ 70\% of the H$_2$ 0$-$0 S(3) line flux in the blueshifted outflow and $\sim$ 40\% in the redshifted outflow is emitted by gas with velocities in excess of 850 km s$^{-1}$. Assuming uniform physical conditions throughout the outflowing material (such that the warm-molecular gas mass directly scales with H$_2$ 0-0 S(3) line flux), we find that a total of 60\% of the outflowing material has a velocity that exceeds the escape velocity of 850 km s$^{-1}$. The warm-molecular outflow traced by H$_2$ 0$-$0, and by extension the cold-molecular outflow which has similar kinematics, may be excavating the buried QSO in the process, depositing the molecular gas in the circumgalactic medium (CGM) of \fzeroeight\ and beyond.  

The large observed velocities of the warm- and cold-molecular gas represent a major challenge for numerical simulations of galactic winds. Cool gas clouds in fast hot winds are shocked and destroyed on time scales generally much shorter than the acceleration time scale needed for this gas to reach distances of 1 kpc and acquire velocities in excess of 1000 \kms\ (let alone 3000 \kms\ before deprojection). This problem may be alleviated if efficient mixing of the stripped gas with the fast hot fluid creates warm gas with the right velocities and on the right spatial scale \citep[e.g.,][]{gronke2018, gronke2020, sparre2020, abruzzo2022, farber2022, fielding2022, gronke2022, gronke2023}. 
In this scenario, the maximum velocities in the warm-molecular gas are expected to be comparable to the highest measured v$_{\mathrm{max}}$ of the warm ionized gas, as found in \fzeroeight\ \citep{rupke2013_gemini}. 

In another attractive scenario \citep[e.g.,][]{faucher2012, zubovas2012, costa2014}, the outflowing warm-H$_2$ material is the end-product of thermal instabilities in a hot wind produced by the initial ``blast wave'' impact of a near-relativistic AGN wind on the ISM. These generic models predict large warm-to-cold H$_2$ mass ratios \citep[close to 100\%;][]{richings2018},
which are several orders of magnitude larger than the value we measure in \fzeroeight. We note, however, that these model predictions refer to 1 Myr after the blast-wave impact and are not scaled to the physical conditions of \fzeroeight\ (e.g., ISM density and temperature), so a fair comparison between the model predictions and the observed velocity field of the outflow and warm-to-cold H$_2$ mass ratios will have to wait until fine-tuned, next-generation simulations become available.

\begin{acknowledgments}

K.Y.D., J.S., and S.V.\ acknowledge partial financial support by NASA for this research through STScI grants No.\  JWST-ERS-01335, JWST-GO-01865, JWST-GO-02547, JWST GO-03869, and JWST GO-05627. M.P.S. acknowledges support under grants RYC2021-033094-I and CNS2023-145506 funded by MCIN/AEI/10.13039/501100011033 and the European Union NextGenerationEU/PRTR. I.G.B. is supported by the Programa Atracci\'on de Talento Investigador ``C\'esar Nombela'' via grant 2023-T1/TEC-29030 funded by the Community of Madrid. Partial funding for open access provided by the UMD Libraries' Open Access Publishing Fund.

\end{acknowledgments}

\vspace{5mm}
\facilities{JWST (MIRI MRS)}

\software{q3dfit (\citealt{q3d2014}, \citealt{q3d2021}), 
Astropy (\citealt{astropy:2013}, \citealt{astropy:2018}, \citealt{astropy:2022}), 
matplotlib \citep{matplotlib2007},
NumPy \citep{numpy2020},
SciPy \citep{SciPy2020}
}


\begin{thebibliography}{}
\expandafter\ifx\csname natexlab\endcsname\relax\def\natexlab#1{#1}\fi
\providecommand{\url}[1]{\href{#1}{#1}}
\providecommand{\dodoi}[1]{doi:~\href{http://doi.org/#1}{\nolinkurl{#1}}}
\providecommand{\doeprint}[1]{\href{http://ascl.net/#1}{\nolinkurl{http://ascl.net/#1}}}
\providecommand{\doarXiv}[1]{\href{https://arxiv.org/abs/#1}{\nolinkurl{https://arxiv.org/abs/#1}}}

\bibitem[{{Abruzzo} {et~al.}(2022){Abruzzo}, {Bryan}, \& {Fielding}}]{abruzzo2022}
{Abruzzo}, M.~W., {Bryan}, G.~L., \& {Fielding}, D.~B. 2022, \apj, 925, 199, \dodoi{10.3847/1538-4357/ac3c48}

\bibitem[{{Argyriou} {et~al.}(2023){Argyriou}, {Glasse}, {Law}, {Labiano}, {{\'A}lvarez-M{\'a}rquez}, {Patapis}, {Kavanagh}, {Gasman}, {Mueller}, {Larson}, {Vandenbussche}, {Glauser}, {Royer}, {Dicken}, {Harkett}, {Sargent}, {Engesser}, {Jones}, {Kendrew}, {Noriega-Crespo}, {Brandl}, {Rieke}, {Wright}, {Lee}, \& {Wells}}]{argyriou2023}
{Argyriou}, I., {Glasse}, A., {Law}, D.~R., {et~al.} 2023, \aap, 675, A111, \dodoi{10.1051/0004-6361/202346489}

\bibitem[{{Astropy Collaboration} {et~al.}(2013){Astropy Collaboration}, {Robitaille}, {Tollerud}, {Greenfield}, {Droettboom}, {Bray}, {Aldcroft}, {Davis}, {Ginsburg}, {Price-Whelan}, {Kerzendorf}, {Conley}, {Crighton}, {Barbary}, {Muna}, {Ferguson}, {Grollier}, {Parikh}, {Nair}, {Unther}, {Deil}, {Woillez}, {Conseil}, {Kramer}, {Turner}, {Singer}, {Fox}, {Weaver}, {Zabalza}, {Edwards}, {Azalee Bostroem}, {Burke}, {Casey}, {Crawford}, {Dencheva}, {Ely}, {Jenness}, {Labrie}, {Lim}, {Pierfederici}, {Pontzen}, {Ptak}, {Refsdal}, {Servillat}, \& {Streicher}}]{astropy:2013}
{Astropy Collaboration}, {Robitaille}, T.~P., {Tollerud}, E.~J., {et~al.} 2013, \aap, 558, A33, \dodoi{10.1051/0004-6361/201322068}

\bibitem[{{Astropy Collaboration} {et~al.}(2018){Astropy Collaboration}, {Price-Whelan}, {Sip{\H{o}}cz}, {G{\"u}nther}, {Lim}, {Crawford}, {Conseil}, {Shupe}, {Craig}, {Dencheva}, {Ginsburg}, {Vand erPlas}, {Bradley}, {P{\'e}rez-Su{\'a}rez}, {de Val-Borro}, {Aldcroft}, {Cruz}, {Robitaille}, {Tollerud}, {Ardelean}, {Babej}, {Bach}, {Bachetti}, {Bakanov}, {Bamford}, {Barentsen}, {Barmby}, {Baumbach}, {Berry}, {Biscani}, {Boquien}, {Bostroem}, {Bouma}, {Brammer}, {Bray}, {Breytenbach}, {Buddelmeijer}, {Burke}, {Calderone}, {Cano Rodr{\'\i}guez}, {Cara}, {Cardoso}, {Cheedella}, {Copin}, {Corrales}, {Crichton}, {D'Avella}, {Deil}, {Depagne}, {Dietrich}, {Donath}, {Droettboom}, {Earl}, {Erben}, {Fabbro}, {Ferreira}, {Finethy}, {Fox}, {Garrison}, {Gibbons}, {Goldstein}, {Gommers}, {Greco}, {Greenfield}, {Groener}, {Grollier}, {Hagen}, {Hirst}, {Homeier}, {Horton}, {Hosseinzadeh}, {Hu}, {Hunkeler}, {Ivezi{\'c}}, {Jain}, {Jenness}, {Kanarek}, {Kendrew}, {Kern}, {Kerzendorf}, {Khvalko}, {King}, {Kirkby}, {Kulkarni},
  {Kumar}, {Lee}, {Lenz}, {Littlefair}, {Ma}, {Macleod}, {Mastropietro}, {McCully}, {Montagnac}, {Morris}, {Mueller}, {Mumford}, {Muna}, {Murphy}, {Nelson}, {Nguyen}, {Ninan}, {N{\"o}the}, {Ogaz}, {Oh}, {Parejko}, {Parley}, {Pascual}, {Patil}, {Patil}, {Plunkett}, {Prochaska}, {Rastogi}, {Reddy Janga}, {Sabater}, {Sakurikar}, {Seifert}, {Sherbert}, {Sherwood-Taylor}, {Shih}, {Sick}, {Silbiger}, {Singanamalla}, {Singer}, {Sladen}, {Sooley}, {Sornarajah}, {Streicher}, {Teuben}, {Thomas}, {Tremblay}, {Turner}, {Terr{\'o}n}, {van Kerkwijk}, {de la Vega}, {Watkins}, {Weaver}, {Whitmore}, {Woillez}, {Zabalza}, \& {Astropy Contributors}}]{astropy:2018}
{Astropy Collaboration}, {Price-Whelan}, A.~M., {Sip{\H{o}}cz}, B.~M., {et~al.} 2018, \aj, 156, 123, \dodoi{10.3847/1538-3881/aabc4f}

\bibitem[{{Astropy Collaboration} {et~al.}(2022){Astropy Collaboration}, {Price-Whelan}, {Lim}, {Earl}, {Starkman}, {Bradley}, {Shupe}, {Patil}, {Corrales}, {Brasseur}, {N{"o}the}, {Donath}, {Tollerud}, {Morris}, {Ginsburg}, {Vaher}, {Weaver}, {Tocknell}, {Jamieson}, {van Kerkwijk}, {Robitaille}, {Merry}, {Bachetti}, {G{"u}nther}, {Aldcroft}, {Alvarado-Montes}, {Archibald}, {B{'o}di}, {Bapat}, {Barentsen}, {Baz{'a}n}, {Biswas}, {Boquien}, {Burke}, {Cara}, {Cara}, {Conroy}, {Conseil}, {Craig}, {Cross}, {Cruz}, {D'Eugenio}, {Dencheva}, {Devillepoix}, {Dietrich}, {Eigenbrot}, {Erben}, {Ferreira}, {Foreman-Mackey}, {Fox}, {Freij}, {Garg}, {Geda}, {Glattly}, {Gondhalekar}, {Gordon}, {Grant}, {Greenfield}, {Groener}, {Guest}, {Gurovich}, {Handberg}, {Hart}, {Hatfield-Dodds}, {Homeier}, {Hosseinzadeh}, {Jenness}, {Jones}, {Joseph}, {Kalmbach}, {Karamehmetoglu}, {Ka{l}uszy{'n}ski}, {Kelley}, {Kern}, {Kerzendorf}, {Koch}, {Kulumani}, {Lee}, {Ly}, {Ma}, {MacBride}, {Maljaars}, {Muna}, {Murphy}, {Norman}, {O'Steen},
  {Oman}, {Pacifici}, {Pascual}, {Pascual-Granado}, {Patil}, {Perren}, {Pickering}, {Rastogi}, {Roulston}, {Ryan}, {Rykoff}, {Sabater}, {Sakurikar}, {Salgado}, {Sanghi}, {Saunders}, {Savchenko}, {Schwardt}, {Seifert-Eckert}, {Shih}, {Jain}, {Shukla}, {Sick}, {Simpson}, {Singanamalla}, {Singer}, {Singhal}, {Sinha}, {Sip{H{o}}cz}, {Spitler}, {Stansby}, {Streicher}, {{{S}}umak}, {Swinbank}, {Taranu}, {Tewary}, {Tremblay}, {Val-Borro}, {Van Kooten}, {Vasovi{'c}}, {Verma}, {de Miranda Cardoso}, {Williams}, {Wilson}, {Winkel}, {Wood-Vasey}, {Xue}, {Yoachim}, {Zhang}, {Zonca}, \& {Astropy Project Contributors}}]{astropy:2022}
{Astropy Collaboration}, {Price-Whelan}, A.~M., {Lim}, P.~L., {et~al.} 2022, \apj, 935, 167, \dodoi{10.3847/1538-4357/ac7c74}

\bibitem[{{Barcos-Mu{\~n}oz} {et~al.}(2017){Barcos-Mu{\~n}oz}, {Leroy}, {Evans}, {Condon}, {Privon}, {Thompson}, {Armus}, {D{\'\i}az-Santos}, {Mazzarella}, {Meier}, {Momjian}, {Murphy}, {Ott}, {Sanders}, {Schinnerer}, {Stierwalt}, {Surace}, \& {Walter}}]{barcos2017}
{Barcos-Mu{\~n}oz}, L., {Leroy}, A.~K., {Evans}, A.~S., {et~al.} 2017, \apj, 843, 117, \dodoi{10.3847/1538-4357/aa789a}

\bibitem[{{Benson} {et~al.}(2003){Benson}, {Bower}, {Frenk}, {Lacey}, {Baugh}, \& {Cole}}]{Benson2003}
{Benson}, A.~J., {Bower}, R.~G., {Frenk}, C.~S., {et~al.} 2003, \apj, 599, 38, \dodoi{10.1086/379160}

\bibitem[{{Bushouse} {et~al.}(2022){Bushouse}, {Eisenhamer}, {Dencheva}, {Davies}, {Greenfield}, {Morrison}, {Hodge}, {Simon}, {Grumm}, {Droettboom}, {Slavich}, {Sosey}, {Pauly}, {Miller}, {Jedrzejewski}, {Hack}, {Davis}, {Crawford}, {Law}, {Gordon}, {Regan}, {Cara}, {MacDonald}, {Bradley}, {Shanahan}, {Jamieson}, {Teodoro}, \& {Williams}}]{Bus2022}
{Bushouse}, H., {Eisenhamer}, J., {Dencheva}, N., {et~al.} 2022, {JWST Calibration Pipeline}, 1.8.2,  Zenodo, \dodoi{10.5281/zenodo.7325378}

\bibitem[{Bushouse {et~al.}(2024)Bushouse, Eisenhamer, Dencheva, Davies, Greenfield, Morrison, Hodge, Simon, Grumm, Droettboom, Slavich, Sosey, Pauly, Miller, Jedrzejewski, Hack, Davis, Crawford, Law, Gordon, Regan, Cara, MacDonald, Bradley, Shanahan, Jamieson, Teodoro, Williams, \& Pena-Guerrero}]{Bus2024}
Bushouse, H., Eisenhamer, J., Dencheva, N., {et~al.} 2024, {JWST} Calibration Pipeline,  Zenodo

\bibitem[{Cicone {et~al.}(2014)Cicone, Maiolino, Sturm, Graciá-Carpio, Feruglio, Neri, Aalto, Davies, Fiore, Fischer, García-Burillo, González-Alfonso, Hailey-Dunsheath, Piconcelli, \& Veilleux}]{cicone2014}
Cicone, C., Maiolino, R., Sturm, E., {et~al.} 2014, Astronomy and Astrophysics, 562, A21, \dodoi{10.1051/0004-6361/201322464}

\bibitem[{{Cole} {et~al.}(2000){Cole}, {Lacey}, {Baugh}, \& {Frenk}}]{cole2000}
{Cole}, S., {Lacey}, C.~G., {Baugh}, C.~M., \& {Frenk}, C.~S. 2000, \mnras, 319, 168, \dodoi{10.1046/j.1365-8711.2000.03879.x}

\bibitem[{{Condon} {et~al.}(1990){Condon}, {Helou}, {Sanders}, \& {Soifer}}]{condon1990}
{Condon}, J.~J., {Helou}, G., {Sanders}, D.~B., \& {Soifer}, B.~T. 1990, \apjs, 73, 359, \dodoi{10.1086/191472}

\bibitem[{{Condon} {et~al.}(1991){Condon}, {Huang}, {Yin}, \& {Thuan}}]{condon1991}
{Condon}, J.~J., {Huang}, Z.~P., {Yin}, Q.~F., \& {Thuan}, T.~X. 1991, \apj, 378, 65, \dodoi{10.1086/170407}

\bibitem[{{Costa} {et~al.}(2014){Costa}, {Sijacki}, \& {Haehnelt}}]{costa2014}
{Costa}, T., {Sijacki}, D., \& {Haehnelt}, M.~G. 2014, \mnras, 444, 2355, \dodoi{10.1093/mnras/stu1632}

\bibitem[{{Croton} {et~al.}(2006){Croton}, {Springel}, {White}, {De Lucia}, {Frenk}, {Gao}, {Jenkins}, {Kauffmann}, {Navarro}, \& {Yoshida}}]{croton2006}
{Croton}, D.~J., {Springel}, V., {White}, S. D.~M., {et~al.} 2006, \mnras, 365, 11, \dodoi{10.1111/j.1365-2966.2005.09675.x}

\bibitem[{{Davies} {et~al.}(2024){Davies}, {Shimizu}, {Pereira-Santaella}, {Alonso-Herrero}, {Audibert}, {Bellocchi}, {Boorman}, {Campbell}, {Cao}, {Combes}, {Delaney}, {D{\'\i}az-Santos}, {Eisenhauer}, {Esparza Arredondo}, {Feuchtgruber}, {F{\"o}rster Schreiber}, {Fuller}, {Gandhi}, {Garc{\'\i}a-Bernete}, {Garc{\'\i}a-Burillo}, {Garc{\'\i}a-Lorenzo}, {Genzel}, {Gillessen}, {Gonz{\'a}lez Mart{\'\i}n}, {Haidar}, {Hermosa Mu{\~n}oz}, {Hicks}, {H{\"o}nig}, {Imanishi}, {Izumi}, {Labiano}, {Leist}, {Levenson}, {Lopez-Rodriguez}, {Lutz}, {Ott}, {Packham}, {Rabien}, {Ramos Almeida}, {Ricci}, {Rigopoulou}, {Rosario}, {Rouan}, {Santos}, {Shangguan}, {Stalevski}, {Sternberg}, {Sturm}, {Tacconi}, {Villar Mart{\'\i}n}, {Ward}, \& {Zhang}}]{davies2024}
{Davies}, R., {Shimizu}, T., {Pereira-Santaella}, M., {et~al.} 2024, \aap, 689, A263, \dodoi{10.1051/0004-6361/202449875}

\bibitem[{{Dekel} \& {Silk}(1986)}]{dekel1986}
{Dekel}, A., \& {Silk}, J. 1986, \apj, 303, 39, \dodoi{10.1086/164050}

\bibitem[{{Dekel} \& {Woo}(2003)}]{dekel2003}
{Dekel}, A., \& {Woo}, J. 2003, \mnras, 344, 1131, \dodoi{10.1046/j.1365-8711.2003.06923.x}

\bibitem[{{Fabian}(2012)}]{fabian2012}
{Fabian}, A.~C. 2012, \araa, 50, 455, \dodoi{10.1146/annurev-astro-081811-125521}

\bibitem[{{Farber} \& {Gronke}(2022)}]{farber2022}
{Farber}, R.~J., \& {Gronke}, M. 2022, \mnras, 510, 551, \dodoi{10.1093/mnras/stab3412}

\bibitem[{{Faucher-Gigu{\`e}re} \& {Quataert}(2012)}]{faucher2012}
{Faucher-Gigu{\`e}re}, C.-A., \& {Quataert}, E. 2012, \mnras, 425, 605, \dodoi{10.1111/j.1365-2966.2012.21512.x}

\bibitem[{{Fielding} \& {Bryan}(2022)}]{fielding2022}
{Fielding}, D.~B., \& {Bryan}, G.~L. 2022, \apj, 924, 82, \dodoi{10.3847/1538-4357/ac2f41}

\bibitem[{González-Alfonso {et~al.}(2017)González-Alfonso, Fischer, Spoon, Stewart, Ashby, Veilleux, Smith, Sturm, Farrah, Falstad, Meléndez, Graciá-Carpio, Janssen, \& Lebouteiller}]{gonzalez-alfonso2017}
González-Alfonso, E., Fischer, J., Spoon, H. W.~W., {et~al.} 2017, The Astrophysical Journal, 836, 11, \dodoi{10.3847/1538-4357/836/1/11}

\bibitem[{{Gronke} \& {Oh}(2018)}]{gronke2018}
{Gronke}, M., \& {Oh}, S.~P. 2018, \mnras, 480, L111, \dodoi{10.1093/mnrasl/sly131}

\bibitem[{{Gronke} \& {Oh}(2020)}]{gronke2020}
---. 2020, \mnras, 494, L27, \dodoi{10.1093/mnrasl/slaa033}

\bibitem[{{Gronke} \& {Oh}(2023)}]{gronke2023}
---. 2023, \mnras, 524, 498, \dodoi{10.1093/mnras/stad1874}

\bibitem[{{Gronke} {et~al.}(2022){Gronke}, {Oh}, {Ji}, \& {Norman}}]{gronke2022}
{Gronke}, M., {Oh}, S.~P., {Ji}, S., \& {Norman}, C. 2022, \mnras, 511, 859, \dodoi{10.1093/mnras/stab3351}

\bibitem[{Harris {et~al.}(2020)Harris, Millman, van~der Walt, Gommers, Virtanen, Cournapeau, Wieser, Taylor, Berg, Smith, Kern, Picus, Hoyer, van Kerkwijk, Brett, Haldane, del R{\'{i}}o, Wiebe, Peterson, G{\'{e}}rard-Marchant, Sheppard, Reddy, Weckesser, Abbasi, Gohlke, \& Oliphant}]{numpy2020}
Harris, C.~R., Millman, K.~J., van~der Walt, S.~J., {et~al.} 2020, Nature, 585, 357, \dodoi{10.1038/s41586-020-2649-2}

\bibitem[{{Heckman} \& {Best}(2014)}]{heckman2014}
{Heckman}, T.~M., \& {Best}, P.~N. 2014, \araa, 52, 589, \dodoi{10.1146/annurev-astro-081913-035722}

\bibitem[{{Heckman} \& {Best}(2023)}]{heckman2023}
---. 2023, Galaxies, 11, 21, \dodoi{10.3390/galaxies11010021}

\bibitem[{{Herrera-Camus} {et~al.}(2020){Herrera-Camus}, {Janssen}, {Sturm}, {Lutz}, {Veilleux}, {Davies}, {Shimizu}, {Gonz{\'a}lez-Alfonso}, {Rupke}, {Tacconi}, {Genzel}, {Cicone}, {Maiolino}, {Contursi}, \& {Graci{\'a}-Carpio}}]{herrera2020}
{Herrera-Camus}, R., {Janssen}, A., {Sturm}, E., {et~al.} 2020, \aap, 635, A47, \dodoi{10.1051/0004-6361/201936434}

\bibitem[{Hunter(2007)}]{matplotlib2007}
Hunter, J.~D. 2007, Computing in Science \& Engineering, 9, 90, \dodoi{10.1109/MCSE.2007.55}

\bibitem[{Kim {et~al.}(2002)Kim, Veilleux, \& Sanders}]{kim2002}
Kim, D.~C., Veilleux, S., \& Sanders, D.~B. 2002, The Astrophysical Journal Supplement Series, 143, 277, \dodoi{10.1086/343843}

\bibitem[{{Pereira-Santaella} {et~al.}(2014){Pereira-Santaella}, {Spinoglio}, {van der Werf}, \& {Piqueras L{\'o}pez}}]{pereira2014}
{Pereira-Santaella}, M., {Spinoglio}, L., {van der Werf}, P.~P., \& {Piqueras L{\'o}pez}, J. 2014, \aap, 566, A49, \dodoi{10.1051/0004-6361/201423430}

\bibitem[{{Richings} \& {Faucher-Gigu{\`e}re}(2018)}]{richings2018}
{Richings}, A.~J., \& {Faucher-Gigu{\`e}re}, C.-A. 2018, \mnras, 474, 3673, \dodoi{10.1093/mnras/stx3014}

\bibitem[{{Rigopoulou} {et~al.}(2002){Rigopoulou}, {Kunze}, {Lutz}, {Genzel}, \& {Moorwood}}]{rigopoulou2002}
{Rigopoulou}, D., {Kunze}, D., {Lutz}, D., {Genzel}, R., \& {Moorwood}, A.~F.~M. 2002, \aap, 389, 374, \dodoi{10.1051/0004-6361:20020607}

\bibitem[{{Roueff} {et~al.}(2019){Roueff}, {Abgrall}, {Czachorowski}, {Pachucki}, {Puchalski}, \& {Komasa}}]{roueff2019}
{Roueff}, E., {Abgrall}, H., {Czachorowski}, P., {et~al.} 2019, \aap, 630, A58, \dodoi{10.1051/0004-6361/201936249}

\bibitem[{{Roussel} {et~al.}(2007){Roussel}, {Helou}, {Hollenbach}, {Draine}, {Smith}, {Armus}, {Schinnerer}, {Walter}, {Engelbracht}, {Thornley}, {Kennicutt}, {Calzetti}, {Dale}, {Murphy}, \& {Bot}}]{roussel2007}
{Roussel}, H., {Helou}, G., {Hollenbach}, D.~J., {et~al.} 2007, \apj, 669, 959, \dodoi{10.1086/521667}

\bibitem[{{Rupke} {et~al.}(2023{\natexlab{a}}){Rupke}, {Wylezalek}, {Zakamska}, {Veilleux}, {Vayner}, {Bertemes}, {Ishikawa}, {Liu}, {Lim}, {Murphree}, {Whitesell}, {McCrory}, \& {Anicetti}}]{q3d_rupke2023}
{Rupke}, D., {Wylezalek}, D., {Zakamska}, N., {et~al.} 2023{\natexlab{a}}, {q3dfit: PSF decomposition and spectral analysis for JWST-IFU spectroscopy}, Astrophysics Source Code Library, record ascl:2310.004.
\newblock \doeprint{2310.004}

\bibitem[{{Rupke}(2014{\natexlab{a}})}]{rupke2014}
{Rupke}, D. S.~N. 2014{\natexlab{a}}, {IFSFIT: Spectral Fitting for Integral Field Spectrographs}, Astrophysics Source Code Library, record ascl:1409.005

\bibitem[{{Rupke}(2014{\natexlab{b}})}]{q3d2014}
---. 2014{\natexlab{b}}, {IFSFIT: Spectral Fitting for Integral Field Spectrographs}.
\newblock \doeprint{1409.005}

\bibitem[{{Rupke} {et~al.}(2021){Rupke}, {Schweitzer}, {Viola}, {Lutz}, {Sturm}, {Spoon}, {Veilleux}, \& {Kim}}]{q3d2021}
{Rupke}, D.~S.~N., {Schweitzer}, M., {Viola}, V., {et~al.} 2021, {QUESTFIT: Fitter for mid-infrared galaxy spectra}.
\newblock \doeprint{2112.002}

\bibitem[{Rupke \& Veilleux(2013{\natexlab{a}})}]{rupke2013_gemini}
Rupke, D. S.~N., \& Veilleux, S. 2013{\natexlab{a}}, The Astrophysical Journal, 768, 75, \dodoi{10.1088/0004-637X/768/1/75}

\bibitem[{Rupke \& Veilleux(2013{\natexlab{b}})}]{rupke2013}
---. 2013{\natexlab{b}}, The Astrophysical Journal, 775, L15, \dodoi{10.1088/2041-8205/775/1/L15}

\bibitem[{Rupke \& Veilleux(2016)}]{rupke2016}
---. 2016, The Astrophysical Journal, 827, L20, \dodoi{10.3847/2041-8205/827/1/L20}

\bibitem[{{Rupke} {et~al.}(2023{\natexlab{b}}){Rupke}, {Wylezalek}, {Zakamska}, {Veilleux}, {Bertemes}, {Ishikawa}, {Liu}, {Sankar}, {Vayner}, {Grace Lim}, {McCrory}, {Murphree}, {Whitesell}, {Shen}, {Liu}, {Barrera-Ballesteros}, {Chen}, {Diachenko}, {Goulding}, {Greene}, {Hainline}, {Hamann}, {Heckman}, {Johnson}, {Lutz}, {L{\"u}tzgendorf}, {Mainieri}, {Nesvadba}, {Ogle}, \& {Sturm}}]{Rupke2023}
{Rupke}, D. S.~N., {Wylezalek}, D., {Zakamska}, N.~L., {et~al.} 2023{\natexlab{b}}, \apjl, 953, L26, \dodoi{10.3847/2041-8213/aced85}

\bibitem[{{Seebeck} {et~al.}(2024){Seebeck}, {Veilleux}, {Liu}, {Rupke}, {Vayner}, {Wylezalek}, {Zakamska}, \& {Bertemes}}]{seebeck2024}
{Seebeck}, J., {Veilleux}, S., {Liu}, W., {et~al.} 2024, arXiv e-prints, arXiv:2409.18086, \dodoi{10.48550/arXiv.2409.18086}

\bibitem[{{Sijacki} {et~al.}(2007){Sijacki}, {Springel}, {Di Matteo}, \& {Hernquist}}]{sijacki2007}
{Sijacki}, D., {Springel}, V., {Di Matteo}, T., \& {Hernquist}, L. 2007, \mnras, 380, 877, \dodoi{10.1111/j.1365-2966.2007.12153.x}

\bibitem[{{Somerville} \& {Primack}(1999)}]{somerville1999}
{Somerville}, R.~S., \& {Primack}, J.~R. 1999, \mnras, 310, 1087, \dodoi{10.1046/j.1365-8711.1999.03032.x}

\bibitem[{{Sparre} {et~al.}(2020){Sparre}, {Pfrommer}, \& {Ehlert}}]{sparre2020}
{Sparre}, M., {Pfrommer}, C., \& {Ehlert}, K. 2020, \mnras, 499, 4261, \dodoi{10.1093/mnras/staa3177}

\bibitem[{{Sturm} {et~al.}(2011){Sturm}, {Gonz{\'a}lez-Alfonso}, {Veilleux}, {Fischer}, {Graci{\'a}-Carpio}, {Hailey-Dunsheath}, {Contursi}, {Poglitsch}, {Sternberg}, {Davies}, {Genzel}, {Lutz}, {Tacconi}, {Verma}, {Maiolino}, \& {de Jong}}]{sturm2011}
{Sturm}, E., {Gonz{\'a}lez-Alfonso}, E., {Veilleux}, S., {et~al.} 2011, \apjl, 733, L16, \dodoi{10.1088/2041-8205/733/1/L16}

\bibitem[{Surace {et~al.}(1998)Surace, Sanders, Vacca, Veilleux, \& Mazzarella}]{surace1998}
Surace, J.~A., Sanders, D.~B., Vacca, W.~D., Veilleux, S., \& Mazzarella, J.~M. 1998, The Astrophysical Journal, 492, 116, \dodoi{10.1086/305028}

\bibitem[{{Vayner} {et~al.}(2023){Vayner}, {Zakamska}, {Ishikawa}, {Sankar}, {Wylezalek}, {Rupke}, {Veilleux}, {Bertemes}, {Barrera-Ballesteros}, {Chen}, {Diachenko}, {Goulding}, {Greene}, {Hainline}, {Hamann}, {Heckman}, {Johnson}, {Grace Lim}, {Liu}, {Lutz}, {L{\"u}tzgendorf}, {Mainieri}, {McCrory}, {Murphree}, {Nesvadba}, {Ogle}, {Sturm}, \& {Whitesell}}]{Vayner2023}
{Vayner}, A., {Zakamska}, N.~L., {Ishikawa}, Y., {et~al.} 2023, \apj, 955, 92, \dodoi{10.3847/1538-4357/ace784}

\bibitem[{{Veilleux} {et~al.}(2005){Veilleux}, {Cecil}, \& {Bland-Hawthorn}}]{veilleux2005}
{Veilleux}, S., {Cecil}, G., \& {Bland-Hawthorn}, J. 2005, \araa, 43, 769, \dodoi{10.1146/annurev.astro.43.072103.150610}

\bibitem[{{Veilleux} {et~al.}(2002){Veilleux}, {Kim}, \& {Sanders}}]{veilleux2002}
{Veilleux}, S., {Kim}, D.~C., \& {Sanders}, D.~B. 2002, \apjs, 143, 315, \dodoi{10.1086/343844}

\bibitem[{{Veilleux} {et~al.}(2020){Veilleux}, {Maiolino}, {Bolatto}, \& {Aalto}}]{veilleux2020}
{Veilleux}, S., {Maiolino}, R., {Bolatto}, A.~D., \& {Aalto}, S. 2020, \aapr, 28, 2, \dodoi{10.1007/s00159-019-0121-9}

\bibitem[{Veilleux {et~al.}(2009)Veilleux, Rupke, Kim, Genzel, Sturm, Lutz, Contursi, Schweitzer, Tacconi, Netzer, Sternberg, Mihos, Baker, Mazzarella, Lord, Sanders, Stockton, Joseph, \& Barnes}]{veilleux2009}
Veilleux, S., Rupke, D. S.~N., Kim, D.~C., {et~al.} 2009, The Astrophysical Journal Supplement Series, 182, 628, \dodoi{10.1088/0067-0049/182/2/628}

\bibitem[{{Veilleux} {et~al.}(2013){Veilleux}, {Mel{\'e}ndez}, {Sturm}, {Gracia-Carpio}, {Fischer}, {Gonz{\'a}lez-Alfonso}, {Contursi}, {Lutz}, {Poglitsch}, {Davies}, {Genzel}, {Tacconi}, {de Jong}, {Sternberg}, {Netzer}, {Hailey-Dunsheath}, {Verma}, {Rupke}, {Maiolino}, {Teng}, \& {Polisensky}}]{veilleux2013}
{Veilleux}, S., {Mel{\'e}ndez}, M., {Sturm}, E., {et~al.} 2013, \apj, 776, 27, \dodoi{10.1088/0004-637X/776/1/27}

\bibitem[{{Veilleux} {et~al.}(2023){Veilleux}, {Liu}, {Vayner}, {Wylezalek}, {Rupke}, {Zakamska}, {Ishikawa}, {Bertemes}, {Barrera-Ballesteros}, {Chen}, {Diachenko}, {Goulding}, {Greene}, {Hainline}, {Hamann}, {Heckman}, {Johnson}, {Grace Lim}, {Lutz}, {L{\"u}tzgendorf}, {Mainieri}, {Maiolino}, {McCrory}, {Murphree}, {Nesvadba}, {Ogle}, {Sankar}, {Sturm}, \& {Whitesell}}]{Veilleux2023}
{Veilleux}, S., {Liu}, W., {Vayner}, A., {et~al.} 2023, \apj, 953, 56, \dodoi{10.3847/1538-4357/ace10f}

\bibitem[{Virtanen {et~al.}(2020)Virtanen, Gommers, Oliphant, Haberland, Reddy, Cournapeau, Burovski, Peterson, Weckesser, Bright, {van der Walt}, Brett, Wilson, Millman, Mayorov, Nelson, Jones, Kern, Larson, Carey, Polat, Feng, Moore, {VanderPlas}, Laxalde, Perktold, Cimrman, Henriksen, Quintero, Harris, Archibald, Ribeiro, Pedregosa, {van Mulbregt}, \& {SciPy 1.0 Contributors}}]{SciPy2020}
Virtanen, P., Gommers, R., Oliphant, T.~E., {et~al.} 2020, Nature Methods, 17, 261, \dodoi{10.1038/s41592-019-0686-2}

\bibitem[{{Wright} {et~al.}(2023){Wright}, {Rieke}, {Glasse}, {Ressler}, {Garc{\'\i}a Mar{\'\i}n}, {Aguilar}, {Alberts}, {{\'A}lvarez-M{\'a}rquez}, {Argyriou}, {Banks}, {Baudoz}, {Boccaletti}, {Bouchet}, {Bouwman}, {Brandl}, {Breda}, {Bright}, {Cale}, {Colina}, {Cossou}, {Coulais}, {Cracraft}, {De Meester}, {Dicken}, {Engesser}, {Etxaluze}, {Fox}, {Friedman}, {Fu}, {Gasman}, {G{\'a}sp{\'a}r}, {Gastaud}, {Geers}, {Glauser}, {Gordon}, {Greene}, {Greve}, {Grundy}, {G{\"u}del}, {Guillard}, {Haderlein}, {Hashimoto}, {Henning}, {Hines}, {Holler}, {Detre}, {Jahromi}, {James}, {Jones}, {Justtanont}, {Kavanagh}, {Kendrew}, {Klaassen}, {Krause}, {Labiano}, {Lagage}, {Lambros}, {Larson}, {Law}, {Lee}, {Libralato}, {Lorenzo Alverez}, {Meixner}, {Morrison}, {Mueller}, {Murray}, {Mycroft}, {Myers}, {Nayak}, {Naylor}, {Nickson}, {Noriega-Crespo}, {{\"O}stlin}, {O'Sullivan}, {Ottens}, {Patapis}, {Penanen}, {Pietraszkiewicz}, {Ray}, {Regan}, {Roteliuk}, {Royer}, {Samara-Ratna}, {Samuelson}, {Sargent}, {Scheithauer},
  {Schneider}, {Schreiber}, {Shaughnessy}, {Sheehan}, {Shivaei}, {Sloan}, {Tamas}, {Teague}, {Temim}, {Tikkanen}, {Tustain}, {van Dishoeck}, {Vandenbussche}, {Weilert}, {Whitehouse}, \& {Wolff}}]{wright2023}
{Wright}, G.~S., {Rieke}, G.~H., {Glasse}, A., {et~al.} 2023, \pasp, 135, 048003, \dodoi{10.1088/1538-3873/acbe66}

\bibitem[{{Wylezalek} {et~al.}(2022){Wylezalek}, {Vayner}, {Rupke}, {Zakamska}, {Veilleux}, {Ishikawa}, {Bertemes}, {Liu}, {Barrera-Ballesteros}, {Chen}, {Goulding}, {Greene}, {Hainline}, {Hamann}, {Heckman}, {Johnson}, {Lutz}, {L{\"u}tzgendorf}, {Mainieri}, {Maiolino}, {Nesvadba}, {Ogle}, \& {Sturm}}]{Wylezalek2022}
{Wylezalek}, D., {Vayner}, A., {Rupke}, D. S.~N., {et~al.} 2022, \apjl, 940, L7, \dodoi{10.3847/2041-8213/ac98c3}

\bibitem[{{Youngblood} {et~al.}(2018){Youngblood}, {France}, {Ginsburg}, {Hoadley}, \& {Bally}}]{youngblood2018}
{Youngblood}, A., {France}, K., {Ginsburg}, A., {Hoadley}, K., \& {Bally}, J. 2018, \apj, 857, 7, \dodoi{10.3847/1538-4357/aab4f4}

\bibitem[{{Zubovas} \& {King}(2012)}]{zubovas2012}
{Zubovas}, K., \& {King}, A. 2012, \apjl, 745, L34, \dodoi{10.1088/2041-8205/745/2/L34}

\end{thebibliography}
\end{document}